# Numerical simulation of time delay interferometry for a LISA-like mission with the simplification of having only one interferometer

S. V. Dhurandhar [a], W.-T. Ni [b,c], and G. Wang [d]

[a] IUCAA, Postbag 4, Ganeshkind, Pune – 4411 007, India   Tel: 91-2025604101

[b] Department of Physics, National Tsing Hua University, No. 101, Kuang Fu II Rd., Hsinchu, Taiwan, 300, ROC   Tel: 886-5715131x33211   Fax: 886-35723052

[c] Shanghai United Center for Astrophysics (SUCA), Shanghai Normal University, 100 Guilin Road, Shanghai, 200234, China   Tel: 86-15335151623

[d] Purple Mountain Observatory, Chinese Academy of Sciences, No. 2, Beijing W. Road, Nanjing, 210008, China   Tel: 86-15002081176

e-mails: sanjeev@iucaa.ernet.in, weitou@gmail.com, gwanggw@gmail.com

**Abstract**

In order to attain the requisite sensitivity for LISA, laser frequency noise must be suppressed below the secondary noises such as the optical path noise, acceleration noise etc. In a previous paper (Dhurandhar et al., Class. Quantum Grav., 27, 135013, 2010), we have found a large family of second-generation analytic solutions of time delay interferometry with one arm dysfunctional, and we also estimated the laser noise due to residual time-delay semi-analytically from orbit perturbations due to Earth. Since other planets and solar-system bodies also perturb the orbits of LISA spacecraft and affect the time delay interferometry (TDI), we simulate the time delay numerically in this paper for all solutions with the generation number $n \leq 3$. We have worked out a set of 3-year optimized mission orbits of LISA spacecraft starting at January 1, 2021 using the CGC2.7 ephemeris framework. We then use this numerical solution to calculate the residual optical path differences in the second-generation solutions of our previous paper, and compare with the semi-analytic error estimate. The accuracy of this calculation is better than 1 cm (or 30 ps). The maximum path length difference, for all configuration calculated, is below 1 m (3 ns). This is well below the limit under which the laser frequency noise is required to be suppressed. The numerical simulation in this paper can be applied to other space-borne interferometers for gravitational wave detection with the simplification of having only one interferometer.

(Submitted February 25, 2011)

1. Introduction

The present gravitational wave (GW) detection activities are focused on the high frequency band (10 Hz – 100 kHz) and the very low frequency band (300 pHz – 100 nHz) using ground interferometers and Pulsar Timing Arrays (PTAs) together with building experimental capacity of space GW detectors for launch in the 2020's to detect low frequency and middle frequency GWs (see, e.g., Ni, 2010, 2012). Ground-based interferometers already have enough sensitivity to set very interesting limits on the strengths of various GW sources in the high frequency band. For example, analysis of the data from a LIGO two-year science run constrains the energy density $\Omega_g$ of the stochastic gravitational-wave background normalized by the critical energy density of the Universe, around 100 Hz, to $6.9 \times 10^{-6}$ and improves on the indirect limit from the Big Bang nucleosynthesis at 100 Hz (The LIGO Scientific Collaboration and the Virgo Collaboration, 2009). Advanced LIGO (The Advanced LIGO Team, 2010), Advanced Virgo (The Advanced Virgo Team, 2010) and LCGT (Kuroda et al., 2010; Nature News 2010) under construction will reach sensitivities that promise a good chance for detecting GWs from binary neutron-star mergers around 2016. PTAs will seek for detection of GWs from supermassive black hole merger background and single events in the very low frequency band around 2020 (Demorest et al., 2009).

In between the high frequency band and the very low frequency band are the middle frequency band (0.1 Hz – 10 Hz) and the low frequency band (100 nHz – 0.1 Hz). Space detectors operate in these bands. Mission concepts under implementation and study are LISA (LISA Study Team, 2000), ASTROD (Bec-Borsenberger et al., 2000; Ni, 2008), Super-ASTROD (Ni, 2009a), ASTROD-GW (Ni, 2009b; Ni et al., 2009), BBO (Crowder and Cornish, 2005) and DECIGO (Kawamura et al., 2006).

LISA—Laser Interferometric Space Antenna—is a proposed ESA-NASA mission which will use coherent laser beams exchanged between three identical spacecraft forming a giant nearly equilateral triangle of side $5 \times 10^6$ km to observe and detect low-frequency cosmic GW (LISA Study Team, 2000). Except possibly in the Fabry-Perot implementation of DECIGO [The requirement on the residual rms motion of the differential arm length is 20 pm (Kawamura et al., 2006). However, if the cavity was kept resonant but the finesses or arm lengths were different or varying, the optical path lengths of each arm would be different and time delay interferometry might still be useful.], all other missions have unequal arm lengths, and laser frequency noise is very serious. One way to suppress it is to use time delay interferometry (TDI) by combining paths to make the two interfering beams to have closely equal optical paths. Time delay interferometry was considered for ASTROD in 1996 (Ni et al., 1997) and has been worked out for LISA much more thoroughly since 1999 (see, e.g., Tinto and Dhurandhar, 2004; and references therein).

In the general orbit model of LISA, the arm lengths vary with time, and the second-generation TDIs are needed for suppressing the laser frequency noise. In this paper, we work out the time delay interferometry *numerically* for LISA with one arm dysfunctional. Obtaining TDI solutions in the general case of time varying arm lengths is a very difficult problem and in our formulation involves the manipulation of non-commutative algebra. In a previous paper (Dhurandhar et al., 2010, henceforth referred to as paper I), we have presented a large family of TDI observables for LISA for which only the data streams from two arms are considered. These apply to the case when one arm of LISA may be dysfunctional. The solutions however are approximate in the sense that the higher order terms involving $\dot{L}^2$ and $\ddot{L}$ or higher are ignored in the calculation, where $L(t)$ is the generic arm length of LISA and the `dot' denotes derivative with respect to time. The importance of the numerical computations is that not only does it test the validity of the TDI observable but also the validity of the approximation. Further, the numerical results are obtained for more realistic orbits of LISA spacecraft that take into account the gravitational effects of most objects in the solar system including several hundred asteroids. This goes beyond paper I, because therein only order of magnitude estimates were used for $\dot{L}^2$ and $\ddot{L}$ based on a semi-analytic model of LISA. We consider optical paths going up to n = 3 (see section 5), that is, successive time-delays up to 23 are considered here. In terms of our formulation, we consider polynomials up to degree 23 in the elementary time-delay operators. There are 14 such TDI observables, which may be deemed sufficient to carry out astrophysical observations. The results of section 5 show that the laser frequency noise is cancelled well within the limit imposed by secondary noises. It should be noted that some of the space missions being discussed in the beginning in this section instead will be characterized by the simplification of having only one interferometer. Thus they are automatically of the type discussed in the paper. [See *note added* at the end of this paper].

In section 2, we review the LISA mission concept, the time-delay interferometry and also the relevant results of paper I. In section 3, we summarize the CGC 2.7 ephemeris framework to be used for optimizing the mission orbit design and to be applied to the second-generation TDI. In section 4, we work out a set of LISA spacecraft orbits to have nearly equal arm lengths and to have minimal line-of-sight Doppler velocity between different pairs of spacecraft. In section 5, we obtain numerical results pertaining to the second-generation TDI mentioned above. In section 6, we conclude this paper with discussion and outlook.

## 2. The technique of TDI in the context of LISA

For laser-interferometric antennas for space detection of GWs, the arm lengths vary according to orbit dynamics. In order to attain the requisite sensitivity, laser frequency noise must be suppressed below the secondary noises such as the optical path noise, acceleration noise etc. There is however redundancy in the data and this can be used to suppress the laser frequency noise by the technique of TDI. The basic principle of TDI is to use two different optical paths but whose optical lengths are equal, or nearly equal, and follow them in opposite directions. This operation suppresses the laser frequency noise if the paths are nearly equal. As argued before in paper I and the previous literature, this amounts to the requirement of the path lengths matching to one part in $10^8$ of LISA arm-length, or within about 50 metres.

In LISA, six data streams arise from the exchange of laser beams between the three spacecraft approximately 5 million km apart. These six streams produce redundancy in the data and they can be combined with appropriate time delays (Armstrong, 2006) to suppress the laser frequency noise. A mathematical foundation for the TDI problem for the static LISA was given in Dhurandhar et al. (2002), where it was shown that the data combinations canceling laser frequency noise formed the *module of syzygies* over the polynomial ring of time-delay operators. For the static LISA, the polynomial ring was in three variables and also commutative. This scheme can be extended in a straightforward way to include the (Sagnac) effect arising from the rotation of LISA, where the up–down optical links are unequal and as a consequence, now one has six variables, but the arm-lengths are still constant in time. Here the polynomial ring is still commutative although over six indeterminates (Nayak and Vinet, 2004). In the general case of time-varying arm-lengths, the polynomial ring is in six variables and also non-commutative. The algebraic problem in the general case has been formulated and discussed in (Dhurandhar, 2009), but the solution to the general problem seems extremely difficult. We still have a linear system that leads to a module (a left module), but it seems difficult to obtain its generators in general. A geometric-combinatorial approach was adopted in Vallisneri (2005) where several solutions were exhibited.

We must envisage the possibility that not all optical links of LISA will be operating at all times for various reasons such as technical failure or even the operating costs. Therefore, it is important to discuss what happens when not all the links operate. Here we look at a specific situation where the data are available from only two arms or four optical links. This should not affect too much the information that can be extracted from the data, because this is essentially a Michelson configuration which is known to be quite sensitive to the GW signal. The practical advantage for this case is that the algebraic problem simplifies considerably and therefore becomes tractable. The problem reduces to one linear constraint on two polynomials. In paper I we have given a family of solutions in a systematic way. The solutions—that is the laser frequency noise is suppressed for these time-delayed data combinations—are approximate in the sense that the $\ddot{L}$ and $\dot{L}^2$ terms are ignored in the calculation, where $L(t)$ is the generic length of the LISA's arm or the optical link and the 'dot' denotes derivative with respect to time. The solutions are based on vanishing commutators and such commutators are enumerated. Further, an algorithm is presented that produces a solution for each such commutator. This analysis can also be applied to other future space detectors like ASTROD-GW (Ni, 2009b; Ni et al., 2009; Wang and Ni, 2010) and BBO (Crowder and Cornish, 2005). However, DECIGO (Kawamura et al., 2006) is an equal arm array and does not need TDI.

Fig. 1 gives a schematic description of LISA where the six links are labelled by $U^i$, $V^i$, $i = 1, 2, 3$, and the optical time-delays by $x$, $y$, $z$ (anti-clockwise) and $l$, $m$, $n$ (clockwise). In our formulation these are *time-delay operators* that act on the data $U^i$, $V^i$ by delaying them by the appropriate optical delays. In the literature it has been shown that for a general combination $X = p_i V^i + q_i U^i$, where $p_i$ and $q_i$ are polynomials in the operators $x$, $y$, $z$, $l$, $m$, $n$, the $p_i$ and $q_i$ satisfy three simultaneous linear equations. However, it must be remembered that the operator multiplication is non-commutative. In the specific case we consider here of one arm being dysfunctional, the three simultaneous equations are essentially reduced to a single non-trivial equation which we need to solve. Assuming without loss of generality the arm connecting SC2 and SC3 to be dysfunctional, we have the non-trivial equation as:

$$p_1 (1 - a) - q_1 (1 - b) = 0, \qquad (1)$$

where $a = lx$ and $b = zn$ are the `round-trip' composite operators. The other non-zero polynomials, namely, $p_3$ and $q_2$ are obtained after solving (1) for $p_1$ and $q_1$:

$$p_3 = - q_1 z, \qquad q_2 = - p_1 l. \qquad (2)$$

The polynomials, $p_2$ and $q_3$ are identically zero. It has been shown in paper I that the commutator $[x_1 x_2 \ldots x_n, y_1 y_2 \ldots y_n]$ is zero in the approximation considered, if $y_1 y_2 \ldots y_n$ is a permutation of $x_1 x_2 \ldots x_n$. Using this, one can construct a host of solutions to (1). The lowest degree solution has already been found before in the literature (Armstrong 2006, Vallisneri 2005). We write it below in our notation:

$$p_1 = 1 - b - ba + ab^2, \qquad q_1 = 1 - a - ab + ba^2. \qquad (3)$$

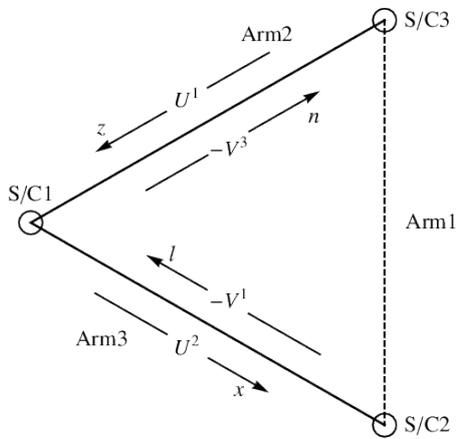

**Figure 1.** The beams and the corresponding time delays are shown schematically in the figure. The functional arms are depicted with a continuous line while the dysfunctional arm is shown with a dashed line.

Then (1) yields the commutator $p_1(1-a) - q_1(1-b) = [ba, ab]$ which is zero in our approximation, because the second term is a permutation of the first. This solution we call n = 1 (one pair of the operators $a$, $b$). In terms of the elementary operators $x$, $l$, $z$, $n$, the $p_1$ and $q_1$ are of degree 6 while $q_2$ and $p_3$ are of degree 7.

As shown in paper I, we can go further and generate more solutions. For n = 2, there are 3 distinct solutions corresponding to three distinct commutators, namely, $[a^2b^2, b^2a^2]$, $[abab, baba]$ and $[ab^2a, ba^2b]$. For n = 3, there are 10 such commutators. The maximum degrees of the polynomials in the elementary operators is $8n - 1$; thus for n = 3, the maximum degree is 23. In section 5 we list all the commutators up to n = 3 and numerically compute the path difference in nanoseconds as a function of time of observation which is taken to be about 1000 days. It is seen from the numerical results that the difference in optical path-lengths and consequently the cancellation of laser frequency noise is well within the requirement.

A related important aspect is the GW response of such TDI observables. The GW response to a TDI observable may be calculated in the simplest way by assuming equal arms (the possible differences in lengths would be sensitive to frequencies outside the LISA bandwidth). A comprehensive and generic treatment of the responses of second-generation TDI observables can be found in Krolak et al. (2004). As remarked, the sensitivity of second-generation TDI observables remains essentially the same as the first-generation ones. The small differences in lengths are important for cancellation/non-cancellation of laser frequency noise and clock noise but not for the GW response. In our case the GW response of all these TDI observables is essentially that of the Michelson.

### 3. CGC ephemeris

In 1998, we started orbit simulation and parameter determination for ASTROD (Chiou and Ni, 2000a, 2000b), and worked out a post-Newtonian ephemeris of the Sun, the major planets and 3 biggest asteroids including the solar quadrupole moment. We term this working ephemeris the CGC 1 (CGC: Center for Gravitation and Cosmology). Using this ephemeris as a deterministic model and adding stochastic terms to simulate noise, we generate simulated ranging data and use Kalman filtering to determine the accuracies of fitted relativistic and solar-system parameters for 1050 days of the ASTROD mission (Chiou and Ni, 2000a, 2000b).

For a better evaluation of the accuracy of $\dot{G}/G$, we need also to monitor the masses of other asteroids. For this, we considered all known 492 asteroids with diameter greater than 65 km to obtain an improved ephemeris framework --- the CGC 2, and calculated the perturbations due to these 492 asteroids on the ASTROD spacecraft (Tang and Ni, 2000, 2002).

In building the CGC ephemeris framework, we use the post-Newtonian barycentric metric and equations of motion as derived in Brumberg (1991) with PPN (Parametrized Post-Newtonian) parameters for solar system bodies. The metric with the gauge parameter α set to zero is

$$ds^2 = [1 - 2\sum_i \frac{m_i}{r_i} + 2\beta(\sum_i \frac{m_i}{r_i})^2 + (4\beta - 2)\sum_i \frac{m_i}{r_i} \sum_{j \neq i} \frac{m_j}{r_{ij}}$$

$$- c^{-2} \sum_i \frac{m_i}{r_i} (2(\gamma+1)\dot{x}_i^2 - r_i \cdot \ddot{x}_i - \frac{1}{r_i^2}(r_i \cdot \dot{x}_i)^2) + \frac{m_1 R_1^2}{r_1^3} J_2 (3(\frac{r_1 \cdot \hat{z}}{r_1})^2 - 1)]c^2 dt^2$$

$$+ 2c^{-1} \sum_i \frac{m_i}{r_i} ((2\gamma+2)\dot{x}_i) \cdot d\mathbf{x} c dt - [1 + 2\gamma \sum_i \frac{m_i}{r_i}](d\mathbf{x})^2 \quad (4)$$

where $r_i = \mathbf{x} - \mathbf{x}_i$, $r_{ij} = \mathbf{x}_i - \mathbf{x}_j$, $m_i = GM_i/c^2$, and $M_i$'s the masses of the bodies with $M_1$ the solar mass (Brumberg, 1991). $J_2$ is the quadrupole moment parameter of the Sun. $\hat{z}$ is the unit vector normal to the ecliptic plane. The associated equations of motion of the N-mass problem derived from the geodesic variational principle of this metric are

$$\ddot{x}_i = -\sum_{j \neq i} \frac{GM_j}{r_{ij}^3} r_{ij} + \sum_{j \neq i} m_j (A_{ij} r_{ij} + B_{ij} \dot{r}_{ij})$$

$$A_{ij} = \frac{\dot{x}_i^2}{r_{ij}^3} - (\gamma+1)\frac{\dot{r}_{ij}^2}{r_{ij}^3} + \frac{3}{2r_{ij}^5}(r_{ij}\dot{x}_j)^2 + G[(2\gamma+2\beta+1)M_i + (2\gamma+2\beta)M_j]\frac{1}{r_{ij}^4}$$

$$+ \sum_{k \neq i,j} GM_k [(2\gamma+2\beta)\frac{1}{r_{ij}^3 r_{ik}} + (2\beta-1)\frac{1}{r_{ij}^3 r_{jk}} + \frac{2(\gamma+1)}{r_{ij}^3 r_{jk}} - (2\gamma+\frac{3}{2})\frac{1}{r_{ik}^3 r_{jk}} - \frac{1}{2r_{jk}^3} \frac{r_{ij} r_{ik}}{r_{ij}^3}]$$

$$B_{ij} = \frac{1}{r_{ij}^3}[(2\gamma+2)(r_{ij}\dot{r}_{ij}) + (r_{ij}\dot{x}_j)] \quad (5)$$

These equations are used to build our computer-integrated ephemeris (with $\gamma = \beta = 1$, $J_2 = 2 \times 10^{-7}$) for nine-planets, the Moon and the Sun. The positions and velocities at the epoch 2005.6.10 0:00 are taken from the DE403 ephemeris. The evolution is solved by using the $4^{th}$-order Runge-Kutta method with a stepsize h =0.01 day. In Chiou and Ni (2000b), the 11-body evolution is extended to 14-body to include the 3 big asteroids --- Ceres, Pallas and Vesta (the CGC 1 ephemeris). Since tilt of the axis of the solar quadrupole moment to the perpendicular of the ecliptic plane is small (7°), in the CGC 1 ephemeris, we have neglected this tilt. In the CGC 2 ephemeris, we have added the perturbations of additional 489 asteroids.

In our previous optimization of ASTROD-GW orbits (Men et al., 2010a, 2010b), we use the CGC 2.5 ephemeris in which only 3 biggest minor planets are taken into account, but the earth's precession and nutation are added; the solar quadratic zonal harmonic and the earth's quadratic to quartic zonal harmonic are also included.

In our recent orbit simulation (Wang and Ni, 2010a) of the ASTROD I proposal to ESA (Braxmaier et al., 2010) and in our study of TDIs (Wang and Ni, 2010b) of ASTROD-GW, we added the perturbation of additional 349 asteroids and called it the CGC 2.7 ephemeris.

In this paper, we use the CGC 2.7 ephemeris to obtain optimized LISA orbits and numerically evaluate TDIs.

We will now consider, as a measure of attainable accuracies, the deviations in calculating the earth orbit. The differences in orbit evolution calculated using CGC 2.7 compared with that of DE405 starting at January 1, 2021 or Earth for 3700 days are shown in Fig. 2 for radial distance, longitude and latitude. The differences in radial distances are less than about 130 m for Earth. The differences for other inner planets (Mercury, Venus and Mars) are smaller.

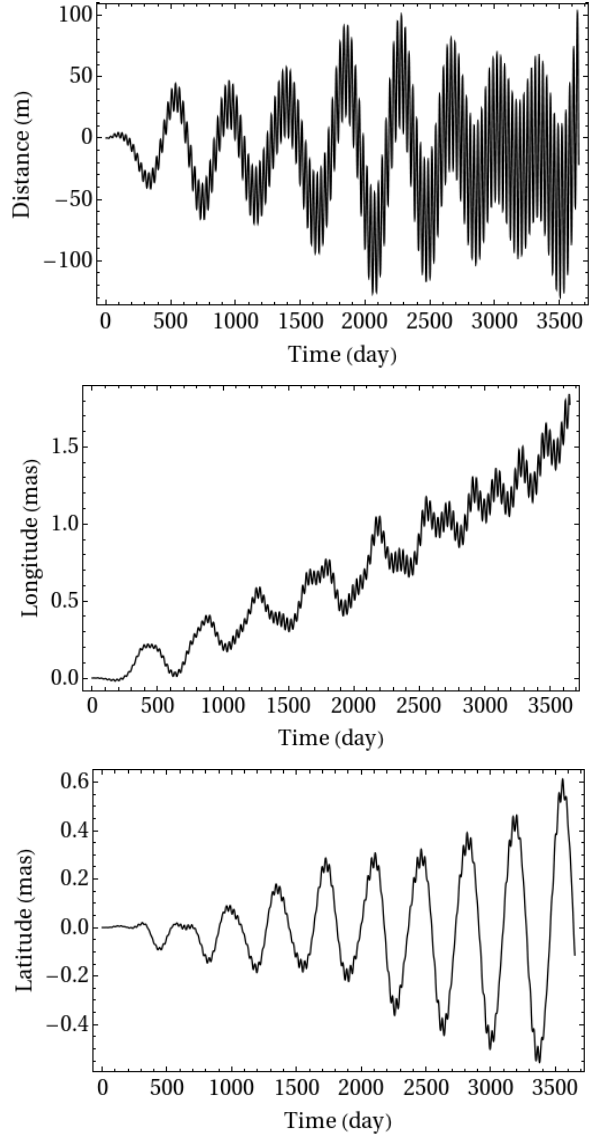

Figure 2: Differences (DE405-CGC2.7) in radial distance, longitude, and latitude for Earth orbits in the DE405 and the CGC2.7 ephemerides in the J2000 Heliocentric Earth mean equatorial coordinate system

## 4. LISA mission orbit optimization

The goal of the LISA mission orbit optimization is to equalize the three arm lengths of the LISA formation and to reduce the relative line-of-sight velocities between three pairs of spacecraft as much as possible. In the previous optimization of Li et al. (2008), the time of start of the science part of the mission is chosen to be 1 January, 2015 (JD2457023.5) and the optimization is for ten years. Since the preparation of the mission will take longer time, in this paper, the time of start of the science part of the mission is taken at 00:00 on January 1st, 2021 (JD2459215.5). This is within 10 years of the optimized orbit of Li et al. (2008). We use the initial conditions of Li et al. and propagate them to January 1, 2021 to obtain a set of initial conditions for the LISA spacecraft at that time (Table I). We then calculated the LISA orbit configuration for 1000 days for working out the optical path length differences of TDIs for comparison with paper I.

Table 1. The initial conditions of 3 spacecraft of LISA in J2000.0 solar-system-barycentric Earth mean equator coordinate system at JD2459215.5, in AU and AU/day.

|  | X/Vx | Y/Vy | Z/Vz |
|---|---|---|---|
| S/C 1 Position | $2.35013223258 \times 10^{-1}$ | $8.97185116074 \times 10^{-1}$ | $4.07383636385 \times 10^{-1}$ |
| S/C 1 Velocity | $-1.65561276145 \times 10^{-2}$ | $3.74073654808 \times 10^{-3}$ | $1.62415066938 \times 10^{-3}$ |
| S/C 2 Position | $2.15130963258 \times 10^{-1}$ | $8.98101216074 \times 10^{-1}$ | $3.80731436385 \times 10^{-1}$ |
| S/C 2 Velocity | $-1.68261446145 \times 10^{-2}$ | $3.56355724808 \times 10^{-3}$ | $1.81794786938 \times 10^{-3}$ |
| S/C 3 Position | $2.47486823258 \times 10^{-1}$ | $8.90689776074 \times 10^{-1}$ | $3.77524966385 \times 10^{-1}$ |
| S/C 3 Velocity | $-1.67592776145 \times 10^{-2}$ | $4.01820804808 \times 10^{-3}$ | $1.46960896938 \times 10^{-3}$ |

The variation of arm lengths and velocities in the line of sight direction are drawn in Figure 3 for 1000 days.

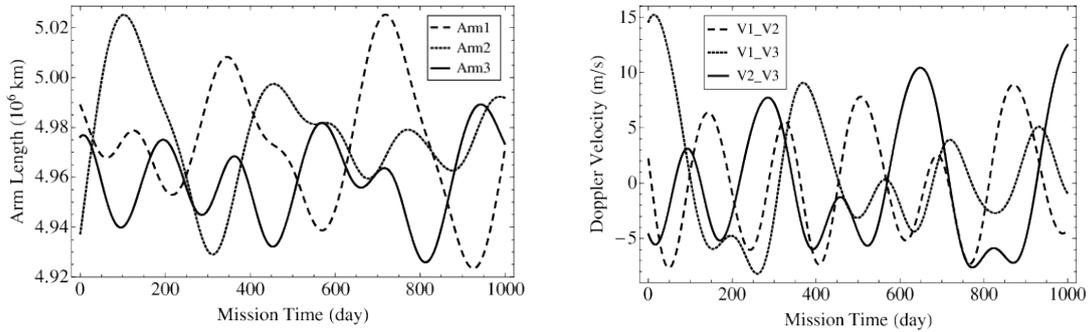

Figure 3 Variations of the arm lengths and the Doppler velocities in 1000 days.

**5. Numerical simulation of second-generation TDI for LISA**

In paper I, we have found a large family of second-generation analytic solutions of time delay interferometry with one arm dysfunctional and estimated the laser noise due to residual time delay semi-analytically from orbit perturbations due to Earth. The method of obtaining these solutions is briefly reviewed in section 2. Here we list these solutions in degree-lexicographic order:

(i) n=1, $[ab, ba] = ab^2a - ba^2b$

(ii) n=2, $[a^2b^2, b^2a^2]$; $[abab, baba]$; $[ab^2a, ba^2b]$

(iii) n=3, $[a^3b^3, b^3a^3]$, $[a^2bab^2, b^2aba^2]$, $[a^2b^2ab, b^2a^2ba]$, $[a^2b^3a, b^2a^3b]$, $[aba^2b^2, bab^2a^2]$, $[ababab, bababa]$, $[abab^2a, baba^2b]$, $[ab^2a^2b, ba^2b^2a]$, $[ab^2aba, ba^2bab]$, $[ab^3a^2, ba^3b^2]$,

where $a = lx$ and $b = zn$ are the `round-trip' composite operators with $l, x, z, n$ as defined in section 2 and figure 1. n = 3 is only an arbitrary upper limit - we could go to higher degrees as well where we may be able to find better choices.

In the numerical calculation, we use the CGC 2.7 ephemeris framework to calculate the path length difference of the two paths for a TDI configuration and plot the difference as function of the epoch of LISA's orbit. We make use of the iteration method (Chiou and Ni, 2000a, 2000b) to calculate the time in the barycentric coordinate system. For the path difference, the accuracy of this calculation is better than 1 cm (or 30 ps) whether we include relativistic light propagation or not. The n = 1, 2, 3 configurations are shown in Figure 4, 5, 6 respectively. We note that all the time differences are below 3 ns (corresponding to the maximum path length difference of 1 m). This is well below the limit under which the laser frequency noise is required to be suppressed and consistent with the semi-analytic results of paper I.

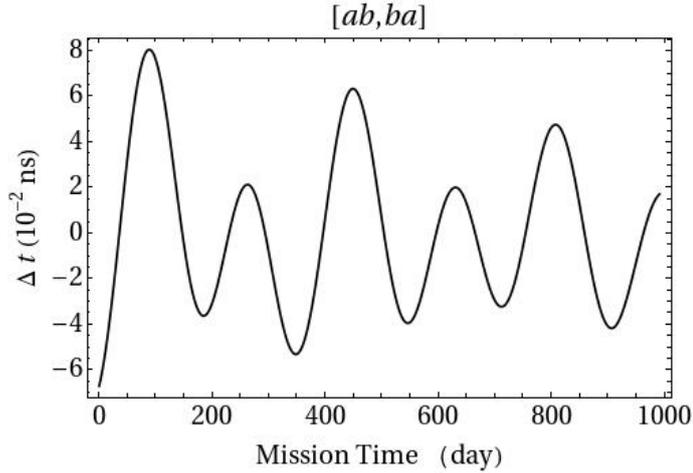

Figure 4: The path length difference (in terms of $\Delta t$ in units of ns [~ 0.3 m]) of two optical paths of n = 1 TDI configuration.

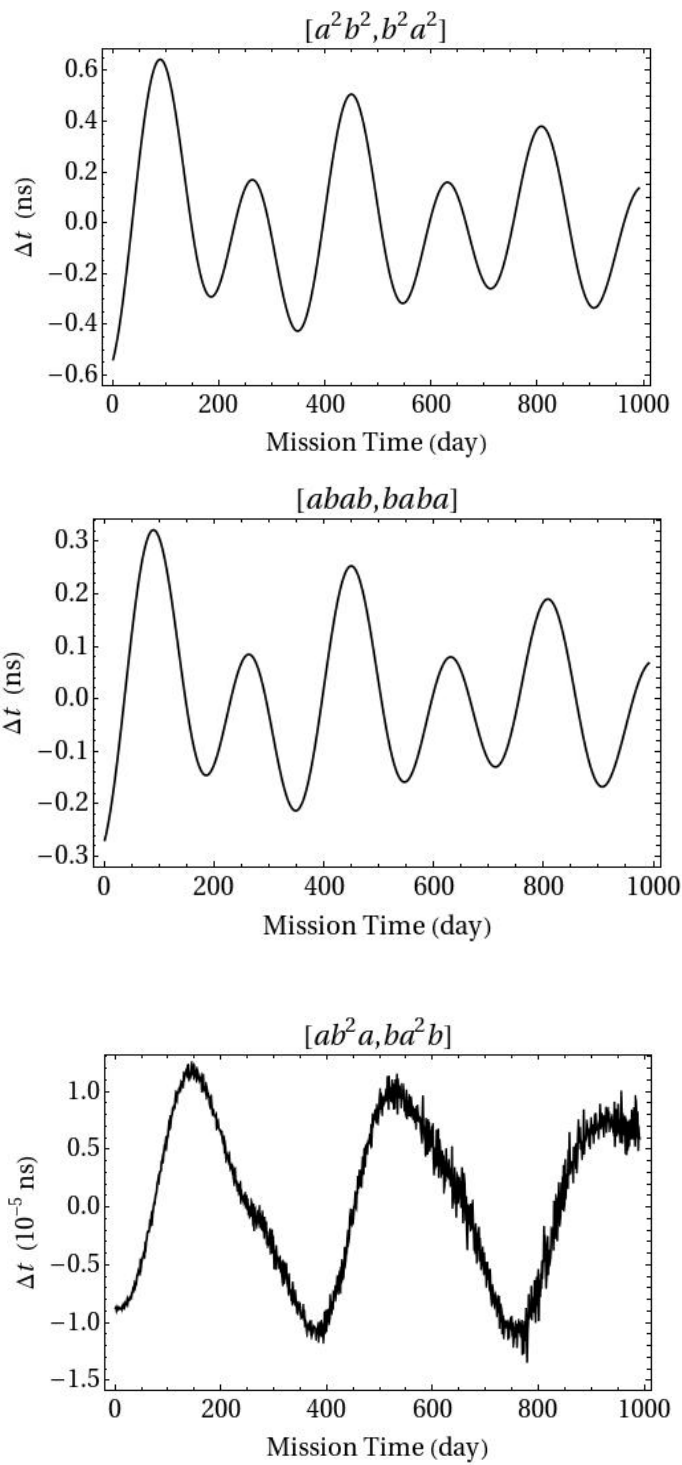

Figure 5: The path length difference (in terms of $\Delta t$ in units of ns [~ 0.3 m]) of two optical paths of three n = 2 TDI configurations.

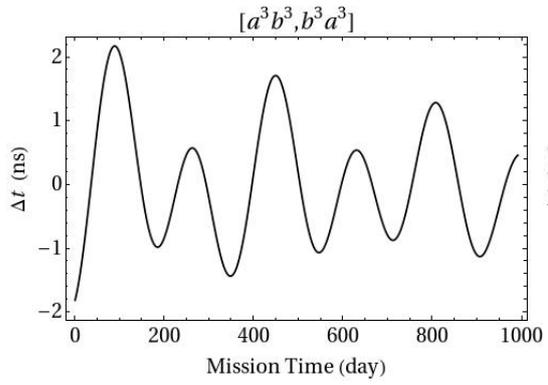
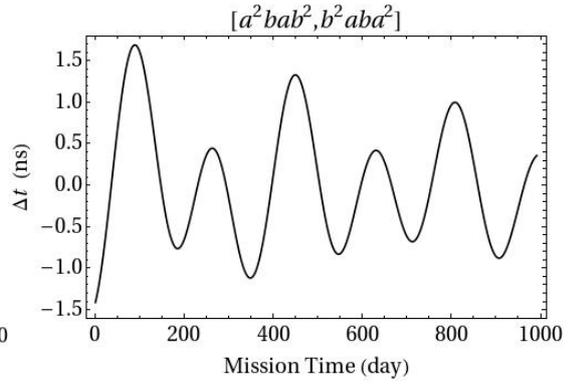
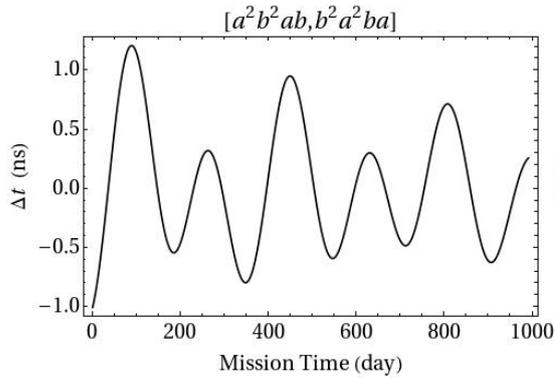
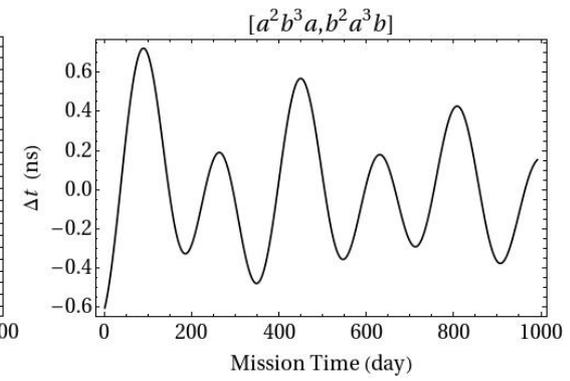
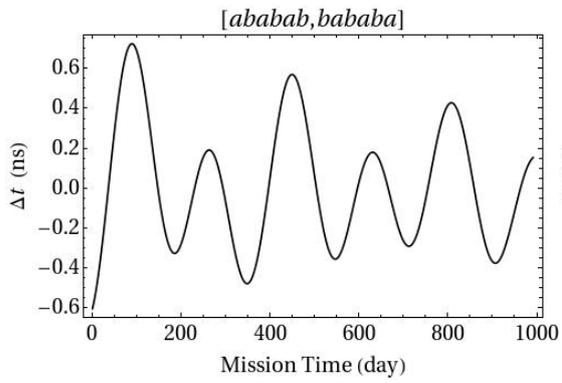
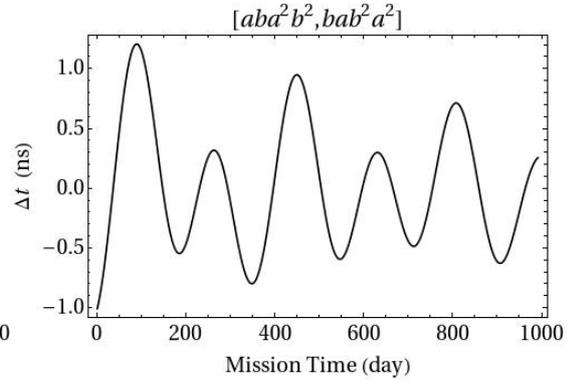
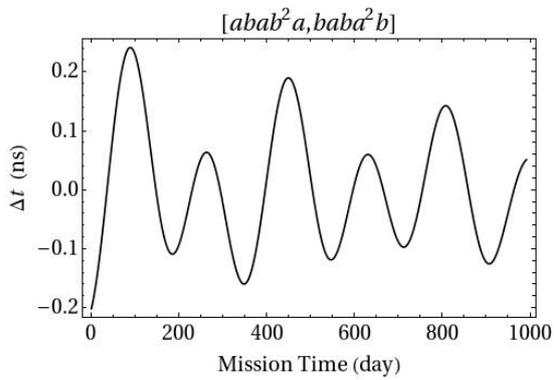
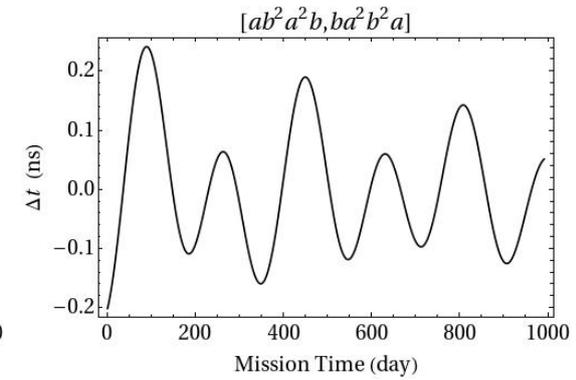

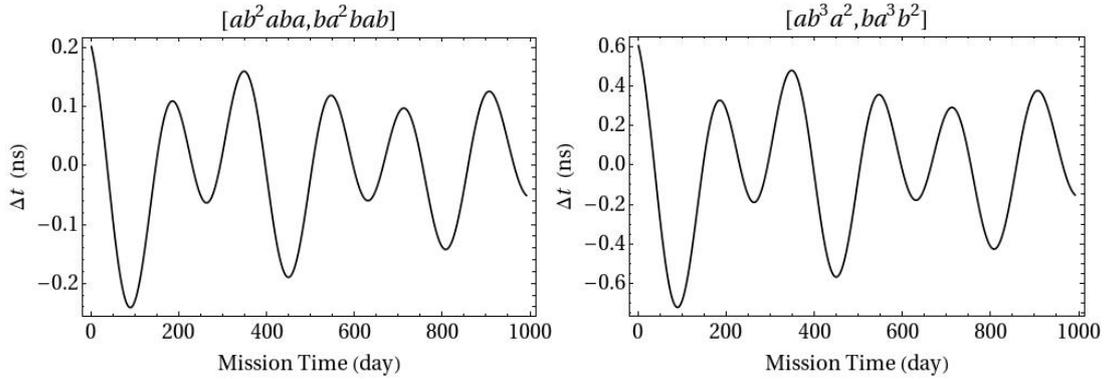

Figure 6: The path length difference (in terms of Δt in units of ns [~ 0.3 m]) of two optical paths of ten n = 3 TDI configurations.

It is interesting to note that for one of the commutators, namely [$ab^2a$, $ba^2b$], the path difference is ~ $10^{-14}$ sec, much below the others. This is result of cancellation of higher order terms in the time derivatives of $L(t)$. Specifically, there is greater symmetry in this combination in which the L double dot terms also cancel out.

## 6. Discussion and Outlook

In the LISA mission for detecting GWs in the frequency range 100 μHz – 1 Hz, the 3 spacecraft range interferometrically with one another with arm lengths of about 5 million kilometers. The previously worked out optimized set of orbits show that the changes of arm length can be less than 10,000 km and the relative Doppler velocities can be less than ±15 m/s for 1000 days. In order to attain the requisite sensitivity for LISA, laser frequency noise must be suppressed below the secondary noises such as the optical path noise, acceleration noise etc. For suppressing laser frequency noise, TDI techniques are used, and in the general case of flexing arms, second-generation TDIs are required. We work out a set of 1000-day optimized mission orbits of LISA spacecraft starting at January 1, 2021, and calculate the residual errors in the second-generation TDI in this paper. The TDIs we consider in this paper are those for which one arm of LISA is dysfunctional and correspond to TDI operator polynomials up to degree 23, that is, 23 consecutive time-delays, which in turn corresponds to n ≤ 3. We have thus examined a total of 14 such TDI solutions which may be deemed sufficient for the purpose of astrophysical observations. All the second-generation TDIs calculated in this paper have optical path differences below 1 m and well satisfy the LISA time offset requirement of 50 m (i.e., 160 ns). The numerical method used here could be readily applied to other (non-dysfunctional) configurations of LISA or LISA-like missions, and we propose to pursue this further.

*Note added*: This work was submitted before it was known that NASA and ESA ceased cooperation. The modified proposal, *e*LISA, taking care of the budget limitation, is just the reduced case we consider in this paper with reduced arm length. The numerical method can be readily applied to this case.

## Acknowledgements

We are grateful to Albrecht Rüdiger for a critical reading of our manuscript and many helpful comments. One of us (WTN) would like to thank IUCAA for providing the superb research and interaction environment during the two visits (January-February, 2009; January-February, 2010) that this collaboration has started and completed.